\documentstyle[12pt]{article}
\newcommand{\be}{\begin{eqnarray}}
\newcommand{\ee}{\end{eqnarray}}

\input{epsfig.sty}

\begin{document}

\begin{figure}[htb]

\epsfxsize=6cm \epsfig{file=logo_INFN.epsf}

\end{figure}

\vspace{-4.75cm}

\Large{\rightline{Sezione ROMA III}}
\large{
\rightline{Via della Vasca Navale 84}
\rightline{I-00146 Roma, Italy}

\vspace{0.6cm}

\rightline{INFN-RM3 98/1}
\rightline{July 1998}
}

\normalsize{}

\vspace{2cm}

\begin{center}

\LARGE{Probing right-handed currents in $B \to K^* \ell^+ \ell^-$
transitions}\\

\vspace{1cm}

\large{D. Melikhov$^{a,b}$, N. Nikitin$^{b}$ and S. Simula$^{c}$}\\

\vspace{0.5cm}

\normalsize{$^a$LPTHE, Universit\'e de Paris XI, B\^atiment 211, 
91405 Orsay Cedex, France\\ $^b$Nuclear Physics Institute, Moscow 
State University, Moscow, 119899, Russia\\ $^c$INFN, Sezione Roma 
III, Via della Vasca Navale 84, I-00146 Roma, Italy}

\end{center}

\vspace{1cm}

\begin{abstract}

We discuss a possibility to probe right-handed weak hadronic currents in 
rare semileptonic $b\to s$ transitions. It is shown that within models 
involving right-handed as well as left-handed quark currents (LR models) 
one can expect a strong enhancement of the right-handed $K^*$ production
in $B \to K^* \ell^+ \ell^-$ decays compared with models including only 
left-handed quark currents (SM, MSSM). Hence an experimental study of the 
transverse asymmetry of the produced $K^*$ mesons provides a clear test 
of the presence of the right-handed quark currents and a possibiltity to 
discriminate between the MSSM and LR extentions of the SM. At the same 
time, MSSM and LR models are found to yield qualitatively the same type 
of deviations from the SM in the forward-backward and the longitudinal 
lepton polarization asymmetries. 

\end{abstract}

\newpage

\pagestyle{plain}

The interest in rare FCNC $B$ decays is motivated to a large extent by the 
fact that these decays provide a possibility to probe the new physics 
effects at comparatively low energies. However, for the experimental study 
of new physics effects it is important to have clear signatures for some 
particular extentions of the SM. Among possible extentions the most popular 
ones are the MSSM and LR models \cite{mssmcmw,mssmhw,lrcm,lrrizzo}. Recently, 
it has been observed \cite{mns2} that e.g. the MSSM extention of the SM can 
be probed by the analysis of the forward-backward ($A_{FB}$) and lepton 
polarization ($P_L$) asymmetries: namely, there are regions in the MSSM 
parameter space which yield qualitatively different behaviors of $A_{FB}$ 
and $P_L$ compared with the SM predictions. Moreover, a possibilty to probe 
the RH currents in $B\to K^*\nu\bar\nu$ decays has been recently pointed 
out in \cite{mns3}. In this letter we discuss a possibility to discriminate 
between the MSSM and LR models by a study of the $q^2$-distributions of the 
transversely polarized $K^*$ mesons produced in the $B \to K^* \ell^+ 
\ell^-$ decays. 

As it is known \cite{lrrizzo}, the parameter space of the LR models is 
rather wide and although the CLEO results on rare radiative decays
provide some restrictions on the values of the LR model parameters, much 
freedom is still left. We report that there are regions in the LR model 
parameter space still allowed by the CLEO data, which yield a strong 
enhancement of the right-handed $K^*$ produced in rare semileptonic (SL) 
$B\to K^*$ transitions. This contrasts to the predictions of other models 
where the RH quark currents are absent (SM, MSSM) and a strong dominance 
of the LH $K^*$ mesons at low $q^2$ is predicted. This property prompts 
that a study of the transverse asymmetry of $K^*$ produced in the $B \to 
K^* \ell^+ \ell^-$ decays can discriminate between the LR models on the 
one hand, and the SM and MSSM on the other. To be more rigorous, a 
difference of the $K^*$ transverse asymmetry from unity would be a clear 
and specific signal of the presence of the RH currents in the Effective 
Hamiltonian. 

We show also that, as fas as other observables, like $A_{FB}$ and $P_L$, 
are concerned, the presence of the RH quark currents in the effective 
Hamiltonian yields generally the same type of deviations from the SM 
predictions as one might expect within the MSSM. 

\section{Effective Hamiltonians and differential distributions}
The effective Hamiltonian for the $b \to s$ transition has the structure 
\cite{gws}:
\begin{equation}
\label{heff}
{\cal H}_{eff} = \frac{G_F}{\sqrt{2}} V_{tb} V_{ts}^\ast\, \sum_i
C_i(\mu) \, O_i(\mu).
\end{equation} 
The operator bases in the SM and MSSM coincide. The operators  which give 
the main contributions are  
\begin{eqnarray}
\label{lbasis}
O_1 &=& \left( \bar{s}_\alpha \gamma^\mu (1-\gamma_5)  b_\alpha \right) 
        \left( \bar{c}_\beta \gamma_\mu (1-\gamma_5) c_\beta \right), 
        \nonumber \\
O_2 &=& \left( \bar{s}_\alpha \gamma^\mu (1-\gamma_5) b_\beta \right)  
        \left( \bar{c}_\beta \gamma_\mu (1-\gamma_5) c_\alpha \right), 
        \nonumber \\
O_{7\gamma} &=& \frac{e}{8\pi^2}\bar{s}_\alpha \sigma_{\mu \nu}
        m_b(\mu)(1+\gamma_5)b_\alpha\ F^{\mu \nu}, \nonumber \\
O_{9V} &=& \frac{e^2}{8\pi^2}(\bar{s}_\alpha \gamma^\mu(1-\gamma_5) b_\alpha) 
        \bar{l} \gamma_\mu l,  \nonumber \\
O_{10A} &=& \frac{e^2}{8\pi^2}(\bar{s}_\alpha \gamma^\mu(1-\gamma_5) b_\alpha)
        \bar{l} \gamma_\mu \gamma_5 l 
\end{eqnarray} 
and the whole difference of the models at large mass scales shows itself 
in the $B$-decays as the difference in the values of the Wilson coefficients 
at the low mass scales. 

The equations for the Wilson coefficients in the SM can be found e.g. in 
\cite{gws}. At the scale $\mu\simeq m_b$ they take the values $C_1(m_b) = 
0.241$, $C_2(m_b) = -1.1$, $C_{7\gamma}(m_b) = 0.312$, $C_{9V}(m_b) = -4.21$, 
$C_{10A}(m_b) = 4.64$. In the MSSM, if SUSY particles have masses above 200 
GeV, the $C^{MSSM}_{9V}$ and $C^{MSSM}_{10A}$ differ from the corresponding 
coefficients in the SM by no more than 10\%. So in all further estimates 
we can safely set $C^{MSSM}_{9V}=C^{SM}_{9V}$ and $C^{MSSM}_{10A} = 
C^{SM}_{10A}$. Fortunately, the difference in the coefficient $C_{7\gamma}$ 
in SM and MSSM might be much more pronounced since the $C^{MSSM}_{7\gamma}$ 
can take values from a broad interval for different regions of the MSSM
parameter space. The experimental results on $B \to K^* \gamma$ and 
$B \to X_s\gamma$ restrict the value $R_{7\gamma}(M_W) = 
C^{MSSM}_{7\gamma}(M_W) / C^{SM}_{7\gamma}(M_W)$ to be in the following 
regions 
\begin{eqnarray}
\label{mssmbounds}
-4.2 < R_{7\gamma} < -2.4, ~~~~  0.4 < R_{7\gamma} < 1.2. 
\end{eqnarray}

In the LR models \cite{lrcm,lrrizzo} the set of the basis operators is 
wider and includes also operators with right-handed quark currents, the 
most important of which are 
\begin{eqnarray}
\label{rbasis}
O^{R}_{1} &=& \left( \bar{s}_\alpha \gamma^\mu (1+\gamma_5)  b_\alpha \right)
        \left( \bar{c}_\beta \gamma_\mu (1+\gamma_5) c_\beta \right), 
        \nonumber \\
O^{R}_{2} &=& \left( \bar{s}_\alpha \gamma^\mu (1+\gamma_5) b_\beta \right)
        \left( \bar{c}_\beta \gamma_\mu (1+\gamma_5) c_\alpha \right), 
        \nonumber \\
O^{R}_{7\gamma} &=& \frac{e}{8\pi^2}\bar{s}_\alpha \sigma_{\mu \nu}
        m_b(\mu)(1-\gamma_5)b_\alpha\ F^{\mu \nu}, \nonumber \\
O^{R}_{9V} &=& \frac{e^2}{8\pi^2}(\bar{s}_\alpha \gamma^\mu(1+\gamma_5) 
        b_\alpha) \bar{l} \gamma_\mu l,  \nonumber \\
O^{R}_{10A} &=& \frac{e^2}{8\pi^2}(\bar{s}_\alpha \gamma^\mu(1+\gamma_5) 
        b_\alpha) \bar{l} \gamma_\mu \gamma_5 l.
\end{eqnarray} 

The result of the analysis of the Wilson coefficients in the LR models 
\cite{lrcm,lrrizzo} shows that in all possible LR model variants the values 
of the Wilson coefficients $C^{R}_{9V}$ and $C^{R}_{10A}$ can be neglected 
compared to the $C^{L}_{9V}$ and $C^{L}_{10A}$ which in turn do not deviate 
considerably from the corresponding SM values. So in the LR case as well as 
in the MSSM, only the difference in $C_{7\gamma}$ is to be taken into 
account. 

The LR model parameter space can be described by the values of the 
right-handed gauge boson mass $M_{W_R}$, the coupling constants of the 
left- and right-handed currents $g_{2L}$ and $g_{2R}$, respectively, the 
mixing angle $\zeta$ and the phase $\beta$ of the gauge boson mass matrix 
\cite{lrrizzo}. The phase $\beta$ is small \cite{lrcm} and can be neglected 
since we are not interested in the small $CP$ violation effects. The Wilson 
coefficients in fact depend only on the combination $\zeta_g = 
(g_{2R}/g_{2L}) \, \zeta$ \cite{lrcm}. Thus the low-energy effective 
Hamiltonian in the LR model actually depends on the values $M_{W_R}$ and 
$\zeta_g$ and the $CKM_R$ matrix elements. To illustrate the possible impact 
of the right-handed currents on the observables in rare SL decays, we use 
one set of the $CKM_R$ parameters from \cite{lrrizzo} denoted as $V_R = 
V_L$ with $M_{W_R} = 1.6 \; TeV$ and the range of $\zeta_g$ determined 
from the CLEO data on $B \to K^* \gamma$. 
 
For the effective Hamilatonian (\ref{heff}) with the operator set given 
by eq. (\ref{lbasis}) and applying the method of Ref. \cite{hag} in the 
case of massless leptons and in the limit $m_s\to0$, one finds for the 
differential distribution $d^4\Gamma / dq^2 d\cos\theta_l d\cos \theta_V 
d\chi$ in a cascade $B \to K^*(\to K \pi) \ell^+ \ell^-$ decay the 
following general expression: 
\begin{eqnarray}
\label{rate}
\frac{d^4\Gamma(B\to K^*(\to K\pi) \ell^+ \ell^-)}{dq^2 d\cos\theta_l 
d\cos\theta_V d\chi}&=&
\frac{3G_F^2}{8(4\pi)^4}\left(\frac{e^2}{8\pi^2}|V^*_{ts}V_{tb}|\right)^2
\frac{\phi^{1/2}M_B}{2}\frac{q^2}{M_B^2}Br(K^*\to K\pi)\\
&&\times \left[ (1-\cos\theta_l)^2\sin^2\theta_V 
\left ( |H^l_{-}|^2+|H^r_{+}|^2 \right )\right . \nonumber \\
&&+(1+\cos\theta_l)^2\sin^2\theta_V
\left (|H^l_{+}|^2+|H^r_{-}|^2 \right ) \nonumber \\
&&+4\sin^2\theta_l\cos^2\theta_V \left (
|H^l_{0}|^2+|H^r_{0}|^2 \right )\nonumber \\
&&-2\sin^2\theta_l\sin^2\theta_V\cos(2\chi)
\left ( {\rm Re}\left (H^l_{+} H^{l*}_{-} \right )+
{\rm Re}\left (H^{r}_{+} H^{r*}_{-} \right )\right )\nonumber \\
&&+2\sin^2\theta_l\sin^2\theta_V\sin(2\chi)
\left ( {\rm Im}\left (H^l_{+} H^{l*}_{-} \right )+
{\rm Im}\left (H^r_{+} H^{r*}_{-} \right )\right )\nonumber \\
&&-2(1-\cos\theta_l)\sin\theta_l\sin (2\theta_V)\cos\chi
\left ( {\rm Re}\left (H^l_{-} H^{l*}_{0} \right )+
        {\rm Re}\left (H^r_{+} H^{r*}_{0}\right )\right )\nonumber \\
&&-2(1-\cos\theta_l)\sin\theta_l\sin(2\theta_V)\sin\chi
\left ( {\rm Im}\left (H^l_{-} H^{l*}_{0} \right )-
        {\rm Im}\left (H^r_{+} H^{r*}_{0}\right )\right )\nonumber \\
&&+2(1+\cos\theta_l)\sin\theta_l\sin(2\theta_V)\cos\chi
\left ( {\rm Re}\left (H^l_{+} H^{l*}_{0}\right )+
        {\rm Re}\left (H^r_{-} H^{r*}_{0}\right )\right )\nonumber \\
&&-2(1+\cos\theta_l)\sin\theta_l\sin(2\theta_V)\sin\chi
\left ( {\rm Im}\left (H^l_{+} H^{l*}_{0}\right )-
        {\rm Im}\left (H^r_{-} H^{r*}_{0}\right )\left.\right )\nonumber
\right]
\end{eqnarray}
where $q = p_B - p_{K^*}$, $\phi = \lambda(1, \hat s, \hat r) = 1 + \hat s^2 
+ \hat r^2 - 2\hat s - 2\hat r - 2\hat s \hat r$, $\hat s = q^2 / M_B^2$ and 
$\hat r = (M_{K^*} / M_B)^2$. The notation of the kinematical variables 
follows the conventional notation of ref. \cite{pdg}. 

The helicity amplitudes $H^{l,r}_{\lambda}$ ($\lambda = 0, \pm$ is the 
$K^*$ meson helicity state) have the following structure in terms of the 
meson transition form factors (see \cite{mns} for their definitions)
\begin{eqnarray}
\label{helicity}
H^{l,r}_{\pm}&=&\bar C^{l,r}f(q^2)-
\frac{C_{7\gamma}}{\hat s}m_b(1-\hat r)B_0(q^2) \nonumber\\
&&\mp\phi^{1/2}\left (\bar C^{l,r}M^2_{B}g(q^2)-
\frac{C_{7\gamma}}{\hat s}m_b g_+(q^2)\right),   \nonumber\\
H^{l,r}_0&=&-\frac{1}{2\sqrt{\hat r\hat s}}
\left [ (1-\hat r-\hat s)\left (\bar C^{l,r}f(q^2) 
-\frac{C_{7\gamma}(m_b)}{\hat s}m_b(1-\hat r)B_0(q^2)
\right )\right. \nonumber\\
&&+\left.\phi\left (\bar C^{l,r}M^2_B a_+(q^2)-
\frac{C_{7\gamma}(m_b)}{\hat s}m_b B_+(q^2)\right)\right], 
\nonumber\\ 
\bar C^{l}&=&\frac12\left(C_{9V}^{eff}(q^2,m_b^2)-C_{10A}(m_b)\right), \qquad
\bar C^{r}=\frac12\left(C_{9V}^{eff}(q^2,m_b^2)+C_{10A}(m_b)\right),
\end{eqnarray}
where the superscripts $l,r$ in $H$ label the helicity structure of the 
corresponding leptonic current. 

The representations (\ref{rate}) and (\ref{helicity}) allow one to obtain 
formulas in various interesting cases making appropriate substitutions. 
The form of such substitutions can be readily obtained from the form of the 
corresponding Effective Hamiltonian, viz.

i. {\bf SL decays, like \boldmath $B \to D^* \ell \nu_{\ell}$ \unboldmath, 
in the SM}: the formula for the decay rate is obtained by substituting
 
\begin{equation}
C_{7\gamma}\to 0,\qquad\bar C^l\to 1,\qquad \bar C^r\to 0,\qquad  
\frac{e^2}{8\pi^2}|V^*_{ts}V_{tb}|\to |V_{bc}|.  
\end{equation}

ii. {\bf Rare decay \boldmath $B\to K^*\nu\bar\nu$ \unboldmath in the SM}:
 
\begin{equation}
C_{7\gamma}\to 0, \qquad 
C_{9V}^{eff}\to \frac{X(x_t)}{\sin^2\theta_W}, \qquad 
C_{10A}\to -\frac{X(x_t)}{\sin^2\theta_W}.  
\end{equation}

iii. {\bf Rare SL decay \boldmath $B\to K^* \ell^+ \ell^-$ \unboldmath in 
the LR models}: if we consider the case of massless leptons and neglect the
$s$-quark mass, then the left- and right-handed parts of both the leptonic 
and the quark currents do not mix with each other and the differential decay 
rate in the LR model, described by the effective Hamiltonian (\ref{heff}) 
with the operator set including (\ref{lbasis}) and (\ref{rbasis}), can be 
obtained by substituting 

\begin{eqnarray}
C_{7\gamma}g_+(q^2)&\to& (C^{L}_{7\gamma}+C^{R}_{7\gamma})g_+(q^2),\nonumber\\
C_{7\gamma}B_0(q^2)&\to& (C^{L}_{7\gamma}-C^{R}_{7\gamma})B_0(q^2),\nonumber\\
C_{7\gamma}B_+(q^2)&\to& (C^{L}_{7\gamma}-C^{R}_{7\gamma})B_+(q^2),\nonumber\\
C_{9V,10A}g(q^2)&\to& (C^{L}_{9V,10A}+C^{R}_{9V,10A})g(q^2),\nonumber\\
C_{9V,10A}f(q^2)&\to& (C^{L}_{9V,10A}-C^{R}_{9V,10A})f(q^2),\nonumber\\
C_{9V,10A}a_\pm(q^2)&\to& (C^{L}_{9V,10A}-C^{R}_{9V,10A})a_\pm(q^2). 
\end{eqnarray}

We are interested only in the nonresonant contribution to the decay rate 
(\ref{rate}), since only the nonresonant part encodes the information on 
the Wilson coefficients. In this case all purely imaginary terms in eq. 
(\ref{rate}) can be neglected. 

The differential distributions of the produced $K^*$ mesons with definite 
helicity takes the form 
\begin{eqnarray}
\label{rh}
\frac{d\Gamma_\lambda}{dq^2}=
\frac{G^2_F}{96\pi^3}\left(\frac{e^2}{8\pi^2}|V^*_{ts}V_{tb}|\right)^2
\frac{\phi^{1/2}M_B}{2}\frac{q^2}{M_B^2}\left[ 
|H^l_\lambda|^2+|H^r_\lambda|^2 \right].
\end{eqnarray}
For the transverse asymmetry defined as 
\begin{eqnarray}
\label{at}
A_T(q^2)=\frac{d\Gamma_-/dq^2-d\Gamma_+/dq^2}
{d\Gamma_-/dq^2 + d\Gamma_+/dq^2}
\end{eqnarray}
one finds the expression 
\begin{eqnarray}
\label{at2}
A_T(q^2)=\frac{2\phi^{1/2} R_T(q^2)}{\phi|G(q^2)|^2+|F(q^2)|^2},
\end{eqnarray}
where 
\begin{eqnarray}
R_T(q^2)&=&Re\left[\left(C^{eff}_{9V}(m_b,q^2)M_B g(q^2)-
\frac{2C_{7\gamma}(q^2)}{\hat s}\frac{m_b}{M_B}g_+(q^2)\right)
\right.\nonumber\\
&\times&\left.\left(C^{eff}_{9V}(m_b,q^2)\frac{f(q^2)}{M_B}-
\frac{2C_{7\gamma}(m_b,q^2)}{\hat s}\frac{m_b}{M_B}
(1-\hat r)B_0(q^2) \right)^*\right] \nonumber\\
&+&|C_{10A}|^2f(q^2)g(q^2)
\end{eqnarray}
and the expressions for G and F can be read off from \cite{mns}.

The angular distribution of the $K$ mesons produced in the subsequent decay 
$K^*\to K\pi$ in the $K^*$ rest frame has the form   
\begin{equation}
\label{ad}
\frac{d\Gamma}{d\cos \theta_V}\sim 1+\alpha\cos^2 \theta_V,
\end{equation}
with
\begin{eqnarray}
\alpha=\frac {\int\limits_{\hat s_{min}}^{\hat s_{max}}d\hat 
s\phi^{\frac{1}{2}}\left (
\left (\frac{(1-\hat r-\hat s)^2}{4\hat r}-\hat s \right )|F(q^2)|^2
-\hat s\phi |G(q^2)|^2
+\frac{\phi^2}{4\hat r}|H_{+}(q^2)|^2-
\frac{\phi (\hat s-1+\hat r)}{2\hat r}R(q^2)\right )}
{\int\limits_{\hat s_{min}}^{\hat s_{max}}d\hat s\phi^{\frac{1}{2}} 
\left ( \hat s\phi |G(q^2)|^2+\hat s|F(q^2)|^2 \right )},
\end{eqnarray} 
where $\hat s_{min}=4m_l^2/M_B^2$ and $\hat s_{max}=(M_B-M_{K^*})^2/M^2_B$. 

\section{Numerical analysis}
In this section we illustrate the possible specific effects which might 
be expected in the LR models due to the presence of the right-handed quark 
currents. In numerical calculations we use the form factors obtained within 
the GI-OGE model\cite{mns}. 

Notice that at large $q^2$ everything is determined by the Wilson coefficients 
$C_{9V}$ and $C_{10A}$ since they are much larger than $C_{7\gamma}$. 
This means that all models (SM, MSSM, LR) give more or less the same results 
for all observables since the deviations in $C_{9V}$ and $C_{10A}$ in all 
extentions of the SM are not large. On the contrary, at small $q^2$ a photon 
pole starts to dominate all observables and the $C_{7\gamma}$ effects are 
enhanced considerably. Since most of the new physics effects lead to 
deviations of $C_{7\gamma}$ from its SM value, different extentions 
of the SM might become distinguishable.

For an illustration of the RH currents influence, we take one of the variants 
of the LR model from \cite{lrrizzo}, namely the extension called $V_L = V_R$. 
In this case the right-handed CKM matrix and $M_{W_R}$ are fixed and the 
freedom of the LR parameter space is reduced to the value of one parameter 
only, namely $\zeta_g$. The allowed range of $\zeta_g$ is constrained by 
the CLEO data \cite{cleo1,cleo2} on rare radiative inclusive and exclusive 
$b \to s \gamma$ transitions  
\begin{eqnarray}
\label{ibr}
Br(B\to X_s\gamma)=\frac{G^2_F\alpha^2_{em}m^5_b}{32\pi^4}|
V^{L*}_{ts}V^L_{tb}|^2\left (|C^L_{7\gamma}(m_b)|^2
+|C^R_{7\gamma}(m_b)|^2\right),
\end{eqnarray}
\begin{eqnarray}
\label{ebr}
Br(B\to K^*\gamma)=\frac{G^2_F\alpha^2_{em}}{32\pi^4}|
V^{L*}_{ts}V^L_{tb}|^2\left (|C^L_{7\gamma}(m_b)|^2
+|C^R_{7\gamma}(m_b)|^2\right)
m^2_b\frac{\left (M^2_B-M^2_{K^*} \right )^3}{M^3_B}
|g_+(0)|^2.
\end{eqnarray}
One finds the allowed region to be
\begin{eqnarray}
-0.02 \leq \zeta_g \leq 0.002.
\end{eqnarray}
Fig. 1 shows the $A_{FB}$ and $P_L$ in the LR model. In the region of small 
$q^2$ the forward-backward asymmetry $A_{FB}$ and the lepton polarization 
asymmetry $P_L$ in MSSM and LR model might be different from the SM but the 
presence of the right-handed quark currents does not add any specific 
effects and one might expect in most favorable case $\zeta_g \simeq -0.02$ 
the same type of deviations from the SM within MSSM and LR models. 

The angular distribution of the secondary $K$ in the cascade decay $B 
\to K^* \ell^+ \ell^- \to K \pi \ell^+ \ell^-$ in the $K^*$ rest frame 
(see eq. (\ref{ad})) turns out to be sensitive to the Wilson coefficients, 
as it is illustrated in Table 1. However, the character of the deviations 
from the SM is similar within the LR and MSSM. So $A_{FB}$, $P_L$ and the 
angular distribution of secondary $K$ mesons can probe the extentions of 
the SM, but they are not sensitive to the specific structure of such 
extentions. 

\begin{table}[h]
\caption{\label{table:alpha}Parameter of the angular distribution of the 
secondary $K$ produced in $B \to K^* \ell^+ \ell^- \to K \pi \ell^+ \ell^-$. 
When calculating $\alpha$ the regions of the resonances have been excluded.}
\vspace{0.25cm}
\centering
\begin{tabular}{|c|c|c|c|c|} \hline
 & SM & \multicolumn{2}{c|} {MSSM} & LR \\
 \hline
 & $R_{7\gamma} = 1$ & $R_{7\gamma} = -4.2 \div -2.4$
 & $R_{7\gamma} = 0.4 \div 1.2$ & $\zeta_g = -0.02 \div 0.002$ \\ 
 \hline
 $\alpha$ & 1.64 & $0.45 \div 1.3$ & $1.6 \div 2.0$ & $0.7 \div 1.8$ \\
 \hline   
\end{tabular}
\end{table}

One might expect that the helicity structure of the Effective Hamiltonian can 
affect the helicity distributions of the final $K^*$ mesons. In fact, 
the distributions of the produced $K^*$ in definite helicity states can be 
considerably affected. Fig. 2 shows the distributions of the right-handed (a) 
and the left-handed (b) $K^*$. One can see that the yield of the right-handed 
$K^*$ in the region of small $q^2$ might be remarkably increased in the LR 
models (see also \cite{mns3}). Such an increase of the right-handed $K^*$ 
mesons in comparison with the predictions of models having suppressed RH 
quark currents, like the SM or the MSSM, can provide a very specific behavior 
of the transverse asymmetry (\ref{at2}), as it is shown clearly in Fig. 3. 
Thus we may conclude that any sizeable difference of $A_T(q^2)$ from unity 
in the region of small $q^2$ would signal the presence of the RH quark 
currents.

Notice that the results presented are not affected significantly by the 
uncertainties in the meson transition form factors: the asymmetyries are 
weakly sensitive to the subtle details of the form factor behavior (see 
the discussion in \cite{mns}) and the right-handed $K^*$ enhancement in 
the LR models far overwhelmes the uncertainties due to the model dependence 
of the form factors. Hence an experimental study of the transversely 
polarized $K^*$ mesons in rare $B \to K^* \ell^+ \ell^-$ decays might 
shed light on the possible presence of the RH currents and their strength.

\vspace{0.5cm} 

\noindent {\bf Acknowledgments.} The authors are grateful to T. Rizzo for 
helpful comments on the LR models.

\newpage

\begin{figure}[htb]
\begin{center}
\begin{tabular}{ccc}
\mbox{\epsfig{file=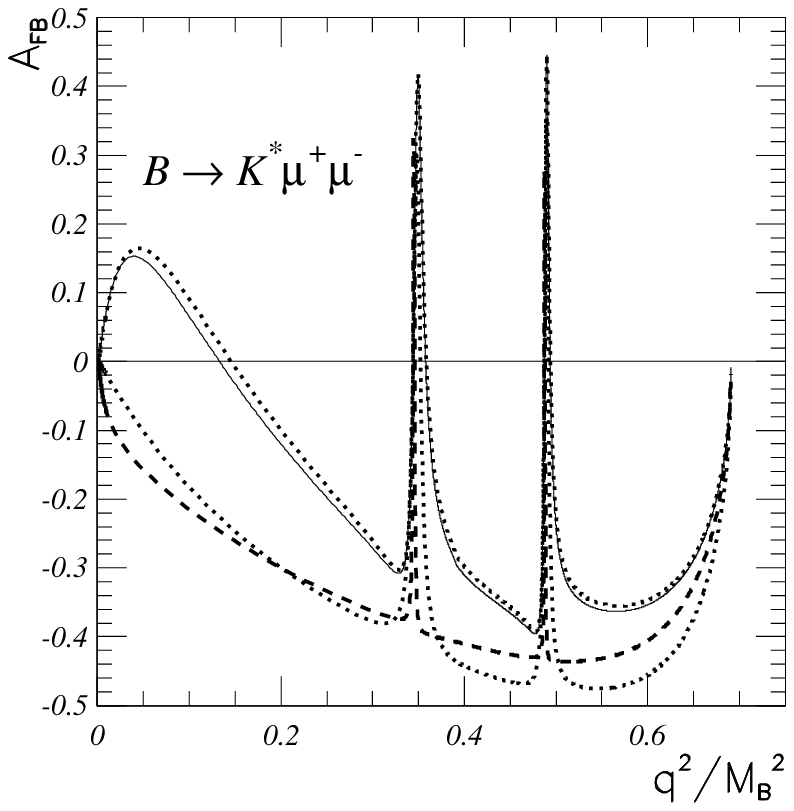,width=7.cm}}
& 
\mbox{\epsfig{file=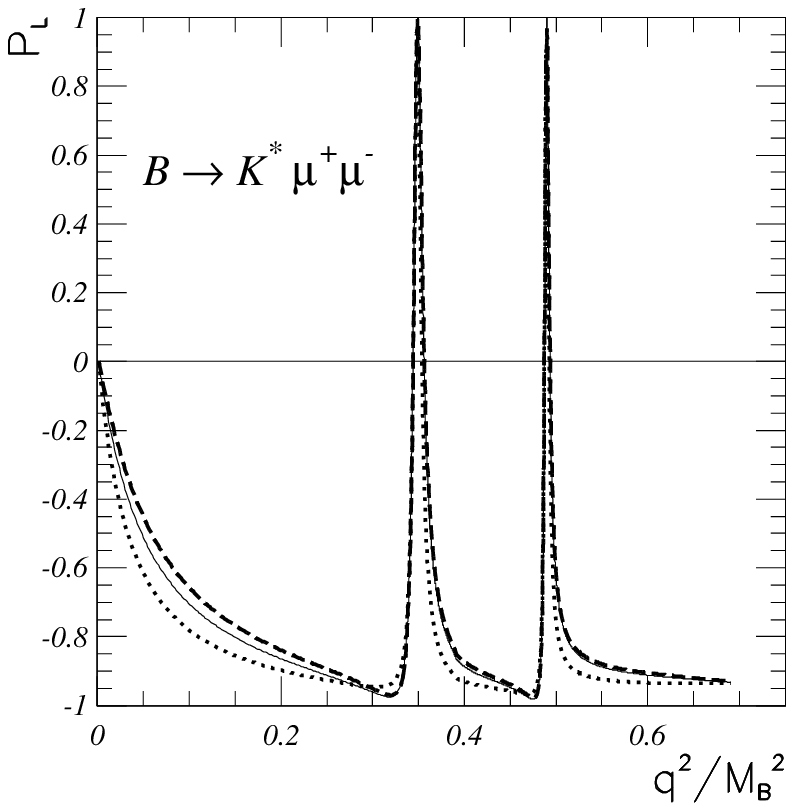,width=7.cm}}
&
\mbox{~~~~~~}
\end{tabular}
\caption{Forward-backward asymmetry $A_{FB}$ and longitudinal lepton 
polarization asymmetry $P_L$ in $B \to K^* \ell^+ \ell^-$ decays: 
solid - SM result, dotted - LR model result ($V_L = V_R$ parameter set), 
lower and upper lines correspond to $\zeta_g = -0.02$ and $\zeta_g = 
0.002$, respectively, dashed line - MSSM with $R_{7\gamma} = -2.4$. 
The values of $P_L$ in the SM and in the LR model with $\zeta_g = 
0.002$ practically coincide and are not distinguishable in the figure. 
\label{fig:1}}
\end{center}
\end{figure}

\begin{figure}[htb]
\begin{center}
\begin{tabular}{ccc}
\mbox{\epsfig{file=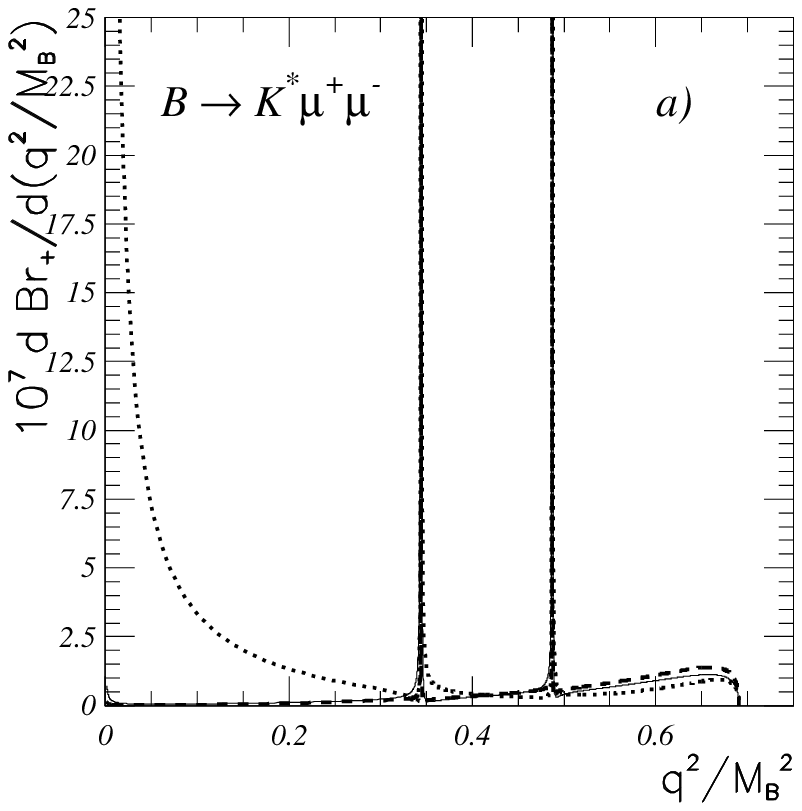,width=7.cm}}
&
\mbox{\epsfig{file=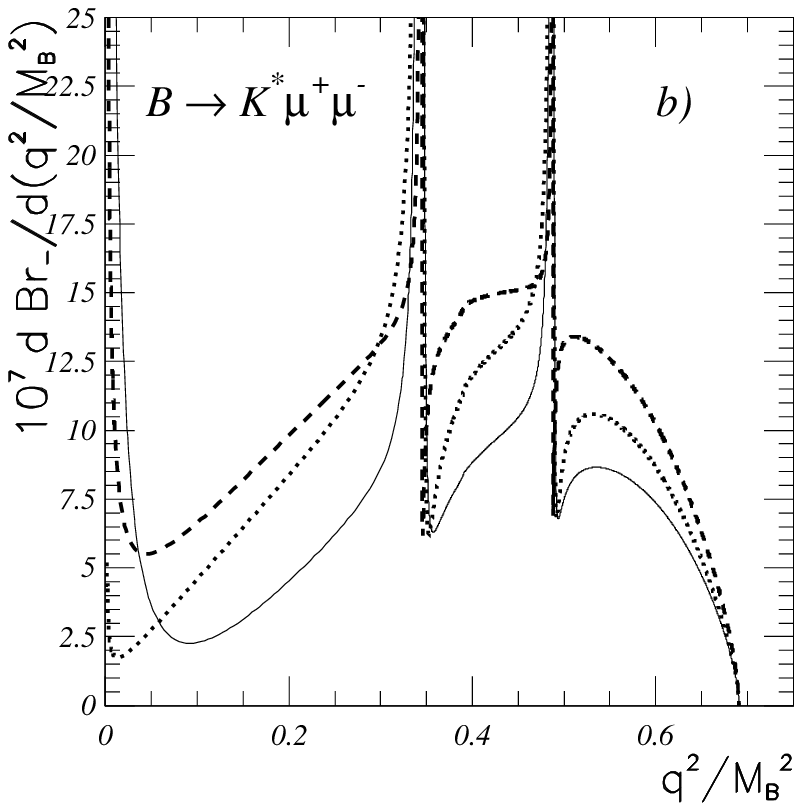,width=7.cm}}
&
\mbox{~~~~~~}
\end{tabular}
\caption{Helicity distributions of $K^*$ produced in $B \to K^* \ell^+ 
\ell^-$ decays: (a) $d\Gamma_+ / dq^2$, (b) $d\Gamma_-/dq^2$: solid - 
SM result, dotted - LR model result ($V_L = V_R$ parameter set) 
corresponding to $\zeta_g = -0.02$, dashed line - MSSM with $R_{7\gamma} 
= -2.4$. 
\label{fig:2}}
\end{center}
\end{figure}

\begin{figure}[htb]
\begin{center}
\mbox{\epsfig{file=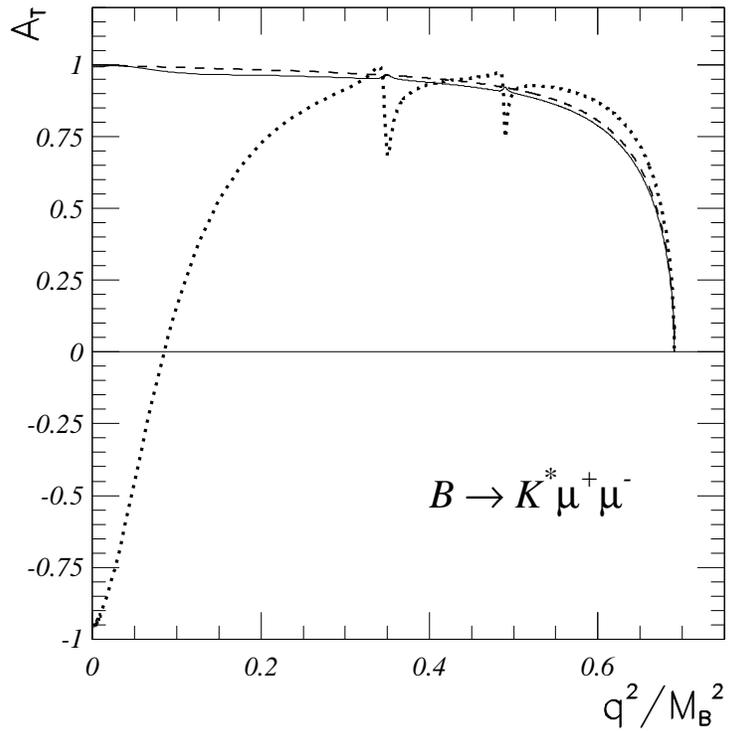,width=12.cm}}
\caption{Differential transverse asymmetry $A_T(q^2)$ of $K^*$ mesons 
produced in $B \to K^* \ell^+ \ell^-$ decays: solid - SM result, dotted 
- LR model result ($V_L = V_R$ parameter set) corresponding to $\zeta_g 
= -0.02$, dashed line - MSSM with $R_{7\gamma} = -2.4$.
\label{fig:3}}
\end{center}
\end{figure}

\end{document}